\documentclass[a4paper]{article}

\usepackage{mathptmx}
\usepackage[utf8]{inputenc} 
\usepackage[T1]{fontenc}    
\usepackage{hyperref}       
\usepackage{url}            
\usepackage{amsfonts}       
\usepackage{amsmath}
\usepackage{nicefrac}       
\usepackage{authblk}        
\usepackage{mathtools}
\usepackage{siunitx}

\usepackage{braket}
\newcommand{\invcm}{cm$^{-1}$}
\newcommand{\um}{\micro\meter}
\newcommand{\acrea}{\hat{a}^{\dagger}}

\newcommand{\ws}{{\omega_s}}
\newcommand{\wi}{{\omega_i}}
\newcommand{\wpump}{{\omega_p}}

\begin{document}

\title{\vspace{-2cm}\bf Mid-IR spectroscopy with NIR grating spectrometers}
\author[1]{Paul Kaufmann}
\author[1]{Helen M Chrzanowski}
\author[2]{Aron Vanselow}
\author[1,3,*]{Sven Ramelow}

\affil[1]{\small Humboldt-Universität zu Berlin, Institut für Physik, Newtonstraße 15, 12489 Berlin, Germany}
\affil[2]{\small L'\'{E}cole Normale Superieure, Labratoire de Physique, 24 rue Lhomond, 75012 Paris, France}
\affil[3]{\small IRIS Adlershof, Humboldt-Universität zu Berlin, Zum Großen Windkanal 6, 12489 Berlin, Germany}
\affil[*]{sven.ramelow@physik.hu-berlin.de}

\date{}
\maketitle

\begin{abstract}
Mid-infrared (mid-IR) spectroscopy is a crucial workhorse for a plethora of analytical applications and is suitable for diverse materials, including gases, polymers or biological tissue. However, this technologically significant wavelength regime between \num{2.5}-\SI{10}{\micro m} suffers from technical limitations primarily related to the large noise in mid-IR detectors and the complexity and cost of bright, broadband mid-IR light sources. Here, using highly non-degenerate, broadband photon pairs from bright spontaneous parametric down-conversion (SPDC) in a nonlinear interferometer, we circumvent these limitations and realise spectroscopy in the mid-IR using only a visible (VIS) solid-state laser and an off-the-shelf, commercial near-infrared (NIR) grating spectrometer. With this proof-of-concept implementation, covering a broad range from \SI{3.2}{\um} to \SI{4.4}{\um} we demonstrate short integration times down to \SI{1}{s} and signal-to-noise ratios above 200 at a spectral resolution from \SI{12}{\invcm} down to \SI{1.5}{\invcm} for longer integration times. Through the analysis of polymer samples and the ambient CO$_2$ in our laboratory, we highlight the potential of this measurement technique for real-world applications.
\end{abstract}

\section*{Introduction}
Spectroscopy in the mid-infrared spectral region (mid-IR) is an indispensable tool in the identification and analysis of a diverse array of materials: from its traditional application to gas \cite{2002Werle} and polymer sensing, its growing use in the life sciences \cite{Fernandez:2005do,Baker:2014ft} and more niche applications, such as igneous rocks \cite{Salisbury:1988uy} and art conservation \cite{Rosi2019}. The diagnostic power of mid-IR spectroscopy stems from the ro-vibrational states of polyatomic molecules, which give rise to strong and specific, fingerprint-like absorption features in the wavelength region between \num{2.5} to \SI{10}{\micro m}.

In contrast to the visible wavelength region, where grating spectrometers are commonly used, the dominant approach for implementing spectroscopy in the mid-IR is Fourier-transform infrared spectroscopy (FTIR) \cite{Griffiths:2007uu}. This is a consequence of mid-IR detector noise, which usually forms the dominant technical limitation but is mitigated in FTIR via its Fellgett advantage \cite{perkins1987fourier}. 
The resulting instrumentation, however, remains complex and bulky, requiring moving parts and often cryogenically cooled detection, rendering it complex to operate and less cost-effective. On the other hand, the resolution of FTIR spectrometers can in principle exceed the maximum practical resolution of grating spectrometers ($\sim$\SI{0.2}{\invcm}) by using sufficiently long scanning distances. This, however, requires increased stability and longer measurement times, such that, in practice, resolutions on the order of \num{1} to \SI{4}{\invcm} are more typical. By comparison, grating spectrometers in the visible and NIR -- absent any moving parts -- can achieve acquisition speeds of \SI{100}{kHz} and more, and are more compact, robust and cost-effective devices. Also, due to the absence of large intrinsic detector noise, there is no Fellgett advantage for VIS-FTIR, which is one of the reasons, why grating spectrometers often dominate in this spectral region.

One promising approach to address the aforementioned challenges of FTIR spectroscopy (speed, robustness, compactness, cost-effectiveness) is to carry out the detection of the mid-IR spectra in the visible wavelength region. This can be achieved, for example, via broadband up-conversion, where a strong pump laser via sum-frequency generation in a nonlinear crystal translates the spectral information from the mid-IR into the visible, where it can be detected with a Si-based detector array \cite{tidemand2016}. In principle such approaches can achieve single-photon sensitivities \cite{Dam:2012hr}. The successful implementation of up-conversion spectroscopy, however, still requires high-power pump lasers for a sufficiently high conversion efficiency and broadband mid-IR light sources, both of which add significant complexity, footprint and cost to the resulting instruments. 

A fundamentally different recently introduced approach for mid-IR spectroscopy relies on nonlinear interferometry \cite{chekhova2016}. It requires neither detectors nor sources operating in the mid-IR. Instead, by exploiting widely non-degenerate correlated photon-pairs generated via spontaneous parametric down-conversion, one can realise sensing and detection at vastly separated wavelengths \cite{kalashnikov2016}. This technique of sensing with undetected photons has also been applied to other modalities of sensing, including imaging \cite{lemos2014,kviatkovsky2020microscopy,buzas2020,paterova2020hyperspectral}, optical coherence tomography \cite{valles2018,paterova2018,vanselow2020oct}, refractometry \cite{paterova2018constants}, and even THz-sensing \cite{kutas2020terahertz,Kutas:21}. 

For the task of spectroscopy, the initial proof of principle demonstration \cite{kalashnikov2016} has seen subsequent improvement \cite{paterova2017}, attaining a spectral resolution of \SI{5}{\invcm} across \SI{120}{\invcm} of bandwidth at \SI{2200}{nm}, albeit with an impractical measurement time (with \SI{30}{s} required for a single fringe). A direct FTIR analog has also been developed \cite{lindner2020fourier} and subsequently improved \cite{Lindner21}, realising a resolution and bandwidth of \SI{0.56}{\invcm} and \SI{700}{\invcm} respectively, with an acquisition time of \SI{900}{s}. 
However, these FTIR implementations are effectively single-pixel approaches, and therefore do not fully exploit a central advantage of the `sensing with undetected photons' paradigm: the use of the detection technologies otherwise absent at the sensing wavelength -- in this case the availability of low-noise, fast ( >\SI{100}{kHz}) and cost-effective kilopixel sensor arrays. 

Here, by directly harnessing the maturity of commercial CCD grating spectrometer technologies, we show mid-IR spectroscopy with undetected photons spanning an \SI{850}{\invcm} large bandwidth and spectral resolutions from \SI{12}{\invcm} to \SI{1.5}{\invcm} with measurement times down to \SI{1}{s}. This is achieved by combining detection with an off-the-shelf grating spectrometer and an ultra-broadband phase-matching condition for the generation of photon pairs \cite{vanselow2019ultra,patent}. To verify our method we demonstrate spectroscopy of polymers (polystyrene and polymethyl methacrylate (PMMA)) as well as gas spectroscopy of atmospheric CO$_2$, with the results yielding excellent agreement with reference FTIR spectra and simulations utilising HITRAN \cite{HITRAN} respectively.

\section*{Experiment and Theory}
\begin{figure*}[t!]
    \centering
    \includegraphics[width=\columnwidth]{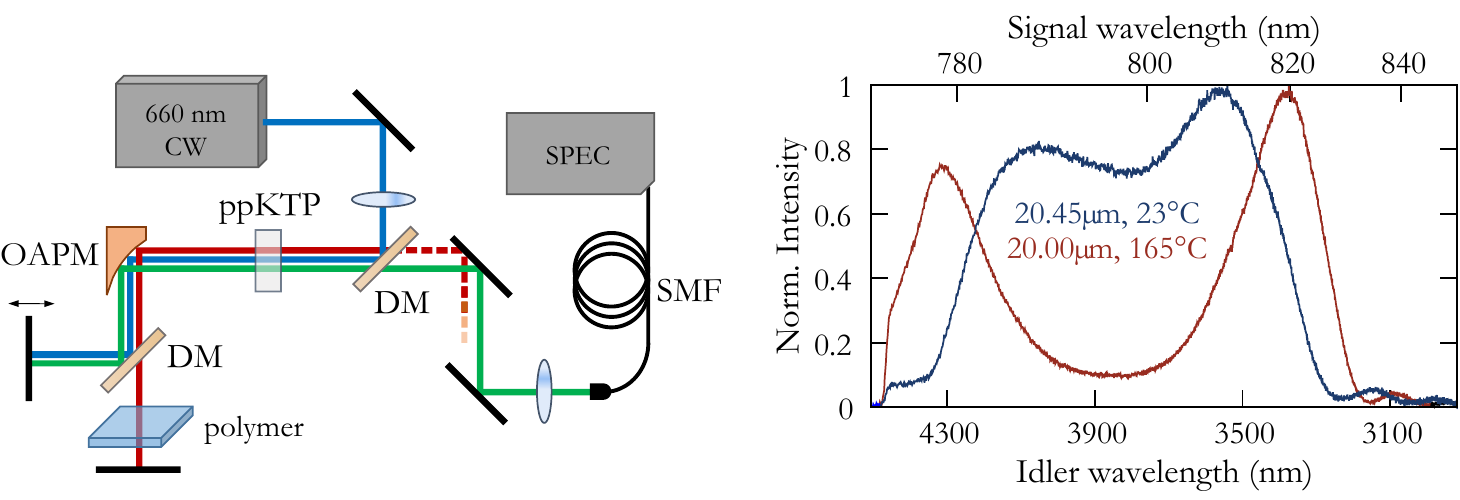}
    \caption{Left:~Nonlinear interferometer setup. A continuous wave laser (blue) pumps the ppKTP crystal to create near-IR signal (green) and mid-IR idler (red) photon pairs. The light is collimated via an off-axis parabolic mirror (OAPM), and mid-IR photons are split from the pump and visible photons via a dichroic mirror (DM). The mid-IR light probes a sample, before being reflected back. After reflecting back all three fields into the crystal, another DM separates the signal photons from the pump, which are subsequently collected into a single mode fiber (SM) and detected by a spectrometer. Right:~Normalized SPDC spectra from the type-0 interactions for the two available poling periods (and crystal temperatures) in our ppKTP crystal. By changing the temperature to fine tune the resulting spectrum we can optimize the photon flux for the spectral region of interest. This allows spectroscopy in the range between 3.2 - \SI{4.4}{\um}.}
    \label{fig:setup}
\end{figure*}

The layout of our nonlinear interferometer resembles a Michelson interferometer where the beamsplitter has been replaced by a nonlinear crystal (see Fig.~\ref{fig:setup}). This crystal can create signal and idler photon pairs either spontaneously, i.e. by an SPDC process, as in our case, or stimulated by other external fields. Pump and signal photons travel in one arm of the interferometer, the other arm carries the idler photon where it can probe a sample. The pump beam propagates through the crystal twice, such that any biphoton created by the annihilation of a pump photon may have originated from \emph{either} the first or second passing of the crystal. Proper alignment renders the two possible processes indistinguishable upon detection. This erasure of `which source' information gives rise to an interference effect, with the visibility of the interference fringes directly related to the indistinguishability of the two processes \cite{mandel1991}.

To understand the theory underlying the experiment presented here, consider an idealised realisation of aforementioned nonlinear interferometer with no absorbing sample in the idler arm. The biphoton state generated by a single pass of the pump light through the crystal is described by
\begin{align}
    \ket{\psi} = \sum_\ws f(\ws) \acrea_\ws \acrea_\wi \ket{0}
           = \sum_\ws f(\ws) \ket{1_\ws, 1_\wi} ,
\end{align}
where $\acrea_{\ws}$ and $\acrea_{\wi}$ are the photon creation operators for signal and idler photons respectively and $f(\ws)$ is the spectrally dependent probability amplitude for the SPDC process, given by $f(\ws) = {\rm sinc}( \Delta k \, L /2)$. Here, $L$ is the crystal length and $\Delta k$ is the wavelength-dependent phase-mismatch $\Delta k = k_p - k_s - k_i - \tfrac{2 \pi}{\Lambda}$ specified for a poling period, $\Lambda$. Owing to the almost monochromatic pump, the idler photon wavelength is tightly correlated to the signal wavelength as a consequence of energy conservation: $\wi = \wpump - \ws$. 
After the second pass of the crystal the biphoton state is given by 
\begin{align}
\ket{\psi}=  \sum_{\omega_s} \frac{1}{\sqrt{2}} f(\omega _s) ( \sqrt{\eta} \ket{1_\ws, 1_\wi} + \sqrt{1-\eta}\ket{1_\ws , 0_\wi} + e^{i\Delta \phi }\ket{1_\ws, 1_\wi} ),
\label{biphotonstate}
\end{align}
where $\eta$ describes the (intensity) transmission of the idler field between the first and second pass of the crystal.
Here, we have explicitly assumed that the first and second SPDC processes are identical and $\eta$ is wavelength independent. The interference between the first and second processes is determined by the total path-difference, $\Delta \phi$, accumulated for all three participating fields between the centre of the first and second crystal \cite{chekhova2016}. This includes a contribution,  $\Delta k L$, from the crystal dispersion itself where $\Delta k$ is the phase mismatch and $L$ is the length of the crystal. Assuming the pump and signal light traverse the same interferometer arm yields
\begin{align}
        \Delta \phi &= 2 \pi \left( \frac{2 (\Delta x + \delta)}{\lambda_p}-\frac{2 (\Delta x + \delta)}{\lambda_s} - \frac{2 (\Delta x )}{\lambda_i} \right) + \Delta k L \nonumber
    \\&= 2 \delta k_i  + \Delta k L,
\end{align}
where $\Delta x$ is the optical path length of the interferometer and $\delta$ accounts any mismatch between the two interferometer arms. Considering a frequency-resolved intensity measurement of the biphoton state of Eqn.~\eqref{biphotonstate} we obtain
\begin{align}
I(\omega_s)= \langle \hat{n}_\ws \rangle=|f(\omega_s)|^2 (1 + \sqrt{\eta} \cos{( 2 \delta k_i  + \Delta k L)}), \label{eq:interference} 
\end{align}
where $k _i = k_p - k_s$. The original biphoton spectrum is modulated by a cosine term, with the frequency increasing with increasing path length mismatch between the fields. By taking the minimum and maximum of Eqn.~\eqref{eq:interference}, we find that the visibility,
\begin{align}
    V = \frac{{\rm max} - {\rm min}}{{\rm max} + {\rm min}} = \sqrt{\eta},
\end{align}
is proportional the square root of the intensity transmission experienced by the idler. Generalising the transmission, $\eta$, to a sample that has a spectrally varying transmission, $\eta ({\omega_i})$, one can see that a measurement of the visibility at the particular signal frequency, $\ws$, allows us to infer the absorption experienced by the idler light at the corresponding wavelength, $\wi = \wpump - \ws$.

Our experimental setup implementation of this measurement scheme is presented in Fig.~\ref{fig:setup}. A single periodically-poled Potassium titanyl phosphate (ppKTP) nonlinear crystal is used in a folded Michelson-style geometry, with this double pass configuration affording significant experimental ease. The crystal itself is purpose engineered to generate very broadband and widely non-degenerate photon pairs \cite{vanselow2019ultra,patent}, providing signal photons in the near-IR for detection and idler photons in the mid-IR for probing the sample. A \SI{660}{nm} continuous wave (CW) laser with tunable output power up to \SI{500}{mW} is focused into the the ppKTP crystal, producing signal and idler fields collinear to the pump, with centre wavelengths of \SI{800}{nm} and \SI{3.8}{\micro m} respectively. All three fields are collimated via an off-axis parabolic mirror and the idler light is subsequently split from the signal and pump light via a dichroic beamsplitter, forming the two distinct arms of the interferometer. The end mirrors of the interferometer arms propagate the light back, recombining and then focusing all three fields back into the crystal. On this second pass, the pump again drives the down-conversion process, closing the nonlinear interferometer. Subsequent to the second pass, we couple the signal (\SI{800}{nm}) light into a single mode fibre for detection by a grating spectrometer, discarding the idler and pump light via another dichroic mirror.

We use an uncooled commercial spectrometer (OceanOptics Maya Pro) with a high sensitivity back thinned charge coupled device (CCD) 2068 pixel array and grating selected for the \SI{800}{nm} region, designed to analyse wavelengths between \num{770} and \SI{860}{nm} with \SI{0.1}{nm} resolution (approximately \SI{1.5}{\invcm}). With our pump wavelength of \SI{660}{nm}, this corresponds to the IR range from \SI{2.8}{\um} to \SI{4.3}{\um}. Our accessible idler wavelengths are limited by absorption in KTP to wavelengths below around \SI{4.3}{\micro m}. However, the important mid-IR band between 3200 and \SI{3600}{nm} used for polymer identification, by their vibrational stretching modes (CH- and OH-groups) is fully accessible with our proof-of-concept setup.

\section*{Results}
\label{sec:results}

\begin{figure*}[t!]
    \centering
    \includegraphics[width=\columnwidth]{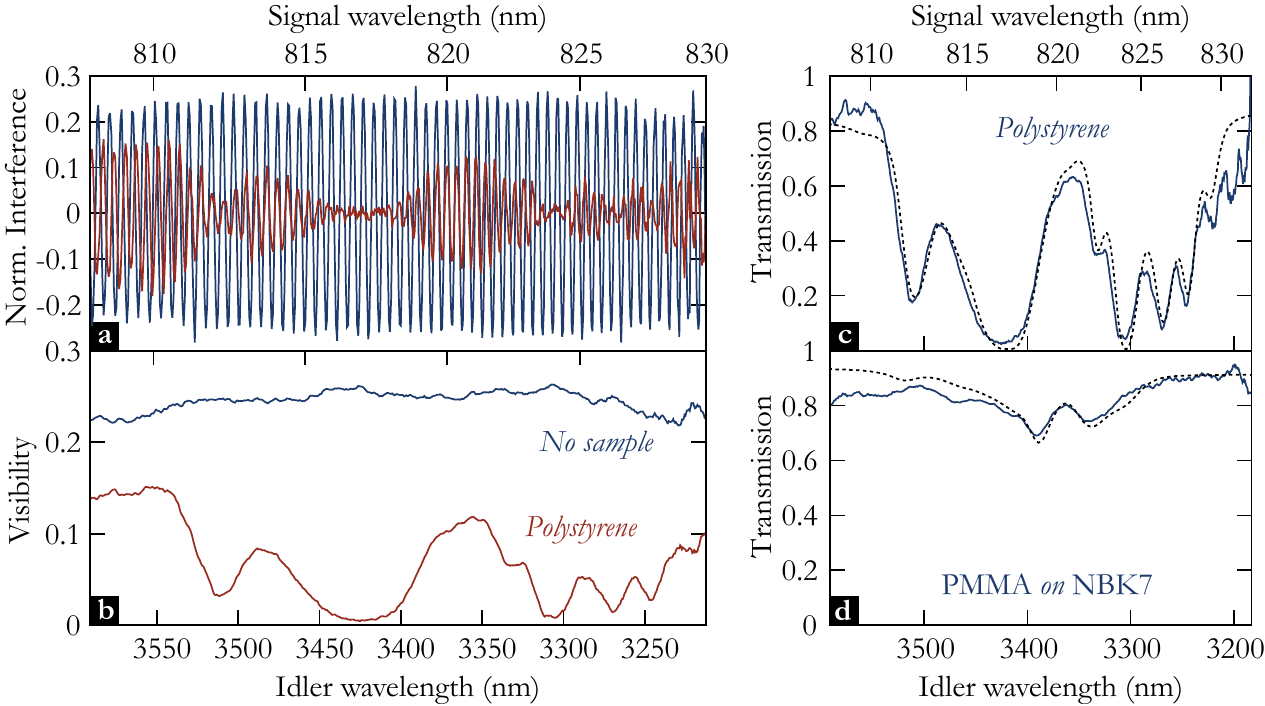}
    \caption{Spectroscopy of polymers via the envelope approach. {\bf a.} The extracted fringe pattern without sample (blue) and with polystyrene testcard (red). {\bf b.} extracted smoothed envelopes of fringe patterns. {\bf c.} Comparison of the resulting transmission spectrum (blue) to FTIR reference (black, dashed). {\bf d.} The same analysis applied to spin-coated PMMA on NBK7 glass slides (blue). The corresponding FTIR reference is also presented (black, dashed).}
    
    \label{fig:EnvelopeResults}
\end{figure*}

With our results we present two distinct approaches to the task of extracting absorption spectra. The first, which we call the `envelope approach', prioritises speed, allowing the acquisition of spectral information with a moderate resolution (comparable to that of commercial FTIR systems) with a single shot. The second 'scanning approach' allows reconstruction of spectra with a resolution determined by that of the near-IR spectrometer, albeit at the expense of longer measurement times and the necessity for a phase-scanning mirror.    

Due to the interferometric nature of the measurement scheme, with the extraction of the absorption spectra from the raw spectrometer data, it is important to consider the simultaneous dependence of the measured signal spectrum on both the absorption and the phase shift simultaneously experienced by the idler field, which by the Kramers-Kronig relations are (non-trivially) related to each other.

\subsubsection*{Envelope approach}
Revisiting the expression for the detected signal spectrum introduced in Eqn.~\eqref{eq:interference},
\begin{align}
    I(\omega_s)=|f(\omega_s)|^2 (1 + \sqrt{\eta} \cos{( 2 \delta k_i + \Delta k L)}), \nonumber
\end{align}
we can recall that our measured spectrum is modulated by a cosine term, with its frequency determined by the displacement $\delta$ introduced between the signal/pump and idler arms. When we insert a sample which has some absorption profile within the spectrum of the idler, the measured (near-IR) signal oscillation will experience an amplitude modulation proportional to the amplitude transmission at the specified wavelength. When double-passed - as is the case here - the instantaneous visibility is directly proportional to the intensity transmission of the sample. Thus, the most straight-forward method and our first approach to extract this absorption information from the modulated spectra is to exploit a Hilbert transform, which allows access to the instantaneous amplitude (or signal envelope) and also the instantaneous phase. 

To obtain a reference, we first take a spectral measurement without a sample, where the interference visibility is largely invariant across the spectrum itself. A sample, in this instance a polystyrene testcard (Perkin \& Elmer), is then placed in the idler arm sufficiently far from the end mirror to ensure any reflections fall outside the post-selected coherence length (as defined by the spectrometer resolution). The polystyrene introduces a spectrally dependent absorption of the idler. Owing to the spectral entanglement, this manifests as amplitude modulation of the signal spectrum. Both recorded spectra (with and without absorption) are normalized by a measurement of the background spectrum, $|f(\omega_s)|^2$, which is recorded by blocking the idler field and therefore suppressing any interference. Dividing out this background spectrum removes the baseband component that arises from the shape of the SPDC spectrum itself. The resulting curves are then centred around zero amplitude (Fig.~\ref{fig:EnvelopeResults}~{\bf a}.) and ideally oscillate at near single frequency. This single frequency will be broadened by dispersion, resulting in a chirp across the spectrum, but can in principle be compensated by both physical and numerical dispersion compensation \cite{wojtkowski2004disp}. 

The two spectra are then processed using a Hilbert transform and subsequent smoothing to extract the visibility at each signal wavelength (Fig.~\ref{fig:EnvelopeResults}~{\bf b}). As both surface reflections and scattering are introduced by the polystyrene testcard, there is a uniform reduction in visibility over the entire spectrum expected for the sample spectra. To correct for this, we re-scale the processed sample spectra accordingly using a region in the spectrum without material absorption. The resulting processed sample spectrum (Fig.~\ref{fig:EnvelopeResults} c.) is in excellent agreement with a reference measurement of the same sample performed with a commercial FTIR unit (Bruker, \SI{4}{\invcm} resolution). Fig.~\ref{fig:EnvelopeResults} d. presents equivalent measurements of polymethyl methacrylate (PMMA). For all the measurements here we used an integration time $t_M$ of \SI{1}{s} per spectrum. 

The resulting envelope approach is single shot and thus enables fast acquisition speed, limited by the acquisition speed of the near-IR spectrometer or by the brightness of the light source. This acquisition speed, however, comes at the expense of maximal resolution. The resolution of the technique is limited by the `carrier' frequency, which is specified by the interferometric displacement. Drawing analogy to signals processing, this frequency should ideally be much faster than the characteristic size of spectroscopic features one would like to resolve. Here, the Nyquist limit imposed by the resolution of the near-IR spectrometer limits this to be greater than \SI{3}{\invcm}, and in practice, even larger still, as several pixels per cycle of the carrier are required to suffer no degradation to the visibility.
Here, estimating the resolution from the size of the smallest resolved features, we achieve a resolution of approximately \SI{12}{\invcm}, or a factor of 8 below the spectrometer resolution. Furthermore, noise can affect the Hilbert transform, as it creates ambiguities in the instantaneous local frequency. Accordingly, use of a filtering around the carrier frequency may be beneficial, but could also introduce some interdependence in the numerical processing and the resolution.

To estimate the signal-to-noise ratio (SNR) in our region of interest between 805 and \SI{830}{nm}, we utilised a moving average to obtain a smoothed visibility, eliminating the bin-to-bin variations we attribute to noise. Taking the standard deviation of the difference between original and averaged visibility, we obtain a SNR of about 200. Note, that this quantity is derived from the `ideal' no-sample case with maximum visibility and is thus only valid for in total, weakly absorbing (trace) gases, whereas for strongly absorbing samples (including the ones tested here) the signal and thus the signal-to-noise ratio will decrease in the more strongly absorbing spectral regions.

\subsubsection*{Phase scanning approach}

To indeed utilise the full resolution of the spectrometer, we introduce a second approach. We scan the position of the mirror in the pump/signal arm and record a single spectrum for every mirror position. Thus, for each individual spectrometer pixel, one can reliably infer the visibility and relative phase-shift for the corresponding idler wavelength. This is in contrast to the aforementioned single shot method, where the visibility is defined locally across several pixels of the near-IR spectrometer.

Again returning to our expression for the measured signal spectrum, consider the measured intensity $I(\omega_s)$ at one spectrometer bin, labelled by $\omega_s$,
\begin{align}
    I(\omega_s)=|f(\omega_s)|^2 (1 + \sqrt{\eta}(\omega_s) \cos{( 2 \delta k_i + \Delta k L)}). \nonumber
\end{align}
Scanning $\delta$ will result in oscillations (interference fringes) for the measured intensity for each pixel with a period set by the corresponding idler wavelength. Given the narrow pump linewidth, the signal wavelength corresponding to the spectrometer bin specifies the idler wavelength, and accordingly the frequency of the oscillation. The measured visibility allows direct inference of the transmission at $\omega_s$.

\begin{figure*}[t!]
  \centering
		\includegraphics[width=\columnwidth]{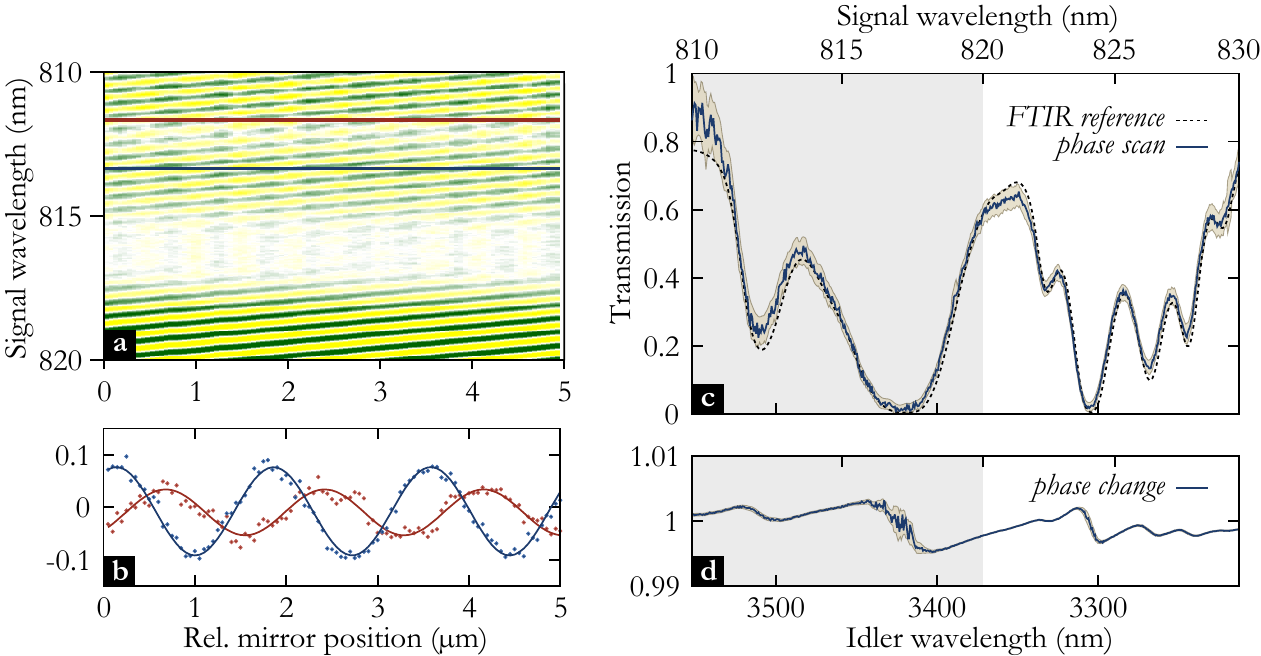}
		\caption{{\bf a.} Surface plot of a phase-scanning measurement. For each mirror position (x-axis) a signal spectrum (y-axis) is obtained. High color contrast corresponds to high visibility of the fringes. The plot shows a limited region of our polystyrene spectrum (from \num{810} to \SI{820}{nm} on the
		spectrometer). {\bf b.} The visibility for each signal wavelength is extracted by fitting a sinusoid to our data (red and blue lines). 
		Red and blue dots correspond to the lines marked in a.
		{\bf c.} The absorption spectrum of the polystyrene testcard (blue) reconstructed via the phase-scanning approach. An FTIR reference measurement (black, dashed) recorded with \SI{4}{\invcm} resolution. The beige shaded region indicates a 95\% confidence interval obtained from the fitting routine. The grey shaded rectangle indicates
		the data range presented in a.}
  \label{fig:Bigplot}
\end{figure*}

Using a nm-precision stage we scan the delay between the two arms of the interferometer and record spectra at equidistant positions (e.g. 100 spectra within \SI{5}{\micro m} of travel). We extract the interference pattern as a function of the stage position (Fig.~\ref{fig:Bigplot}a) and after correcting for the background spectrum, $|f(\ws)|^2$, the fringes at each spectrometer bin (Fig.~\ref{fig:Bigplot}b) are fitted with a cosine:
\begin{eqnarray}
		y = A \cos( 2 k_i x +\phi ) + c.
\end{eqnarray}
The visibility $ A $, the phase $\phi $ and a residual constant offset $c$ are free parameters, while $k_i$ is fixed by the spectrometer calibration. By recording the same procedure without a sample, we also acquire a reference visibility $A_0$ for each bin. As in our envelope approach, we rescale the sample visibility to account for additional losses occurring from reflections and scattering introduced by the sample ($A \rightarrow A'$). The spectral transmission $T$ is then obtained by $T = A'/A_0$.

The outcome of this procedure is shown in Fig.~\ref{fig:Bigplot}. Our results for the polystyrene testcard (Perkin \& Elmer) agree well with a reference measurement performed on the same testcard with an FTIR (Bruker) at \SI{4}{\invcm} resolution (Fig.~\ref{fig:Bigplot}~{\bf c}). We also acquire the phase response from the fitting. The confidence intervals (95\%) shown in the plots are derived from the fitting routine and amount to an error of approximately 5\%. 

Finally, to verify and practically apply the high resolution of our phase scanning approach, we turn our attention to existing absorption of atmospheric carbon dioxide (CO$_2$) within our spectra. Even though the current phase-matching of our crystal accesses idler wavelengths up to \SI{4.5}{\micro m}, spectroscopy beyond \SI{4.1}{\micro m} is largely inaccessible due to a strong CO$_2$ absorption window between \num{4.15}-\SI{4.35}{\micro m} (\num{2300}-\SI{2400}{\invcm}).
Gas spectroscopy, however, offers access to features substantially narrower than those typical of samples in the solid phase, providing us an opportunity to directly verify the resolution of our technique. As a reference measurement without the CO$_2$ is challenging without evacuating the setup or lab, we instead took a modified approach to extract the absorption spectrum of CO$_2$: changing the optical path length of the interferometer and therefore effectively varying the absorption experienced by the idler. At 5 path lengths evenly spaced across \SI{40}{cm} the aforementioned phase scanning technique was implemented and the signal visibility was reconstructed. Then, utilising a linear fit with the known absorption path length and Beer's law, the absorption coefficient versus idler wavelength was reconstructed. 

\begin{figure*}[t!]
  \centering
  \includegraphics[width=\columnwidth]{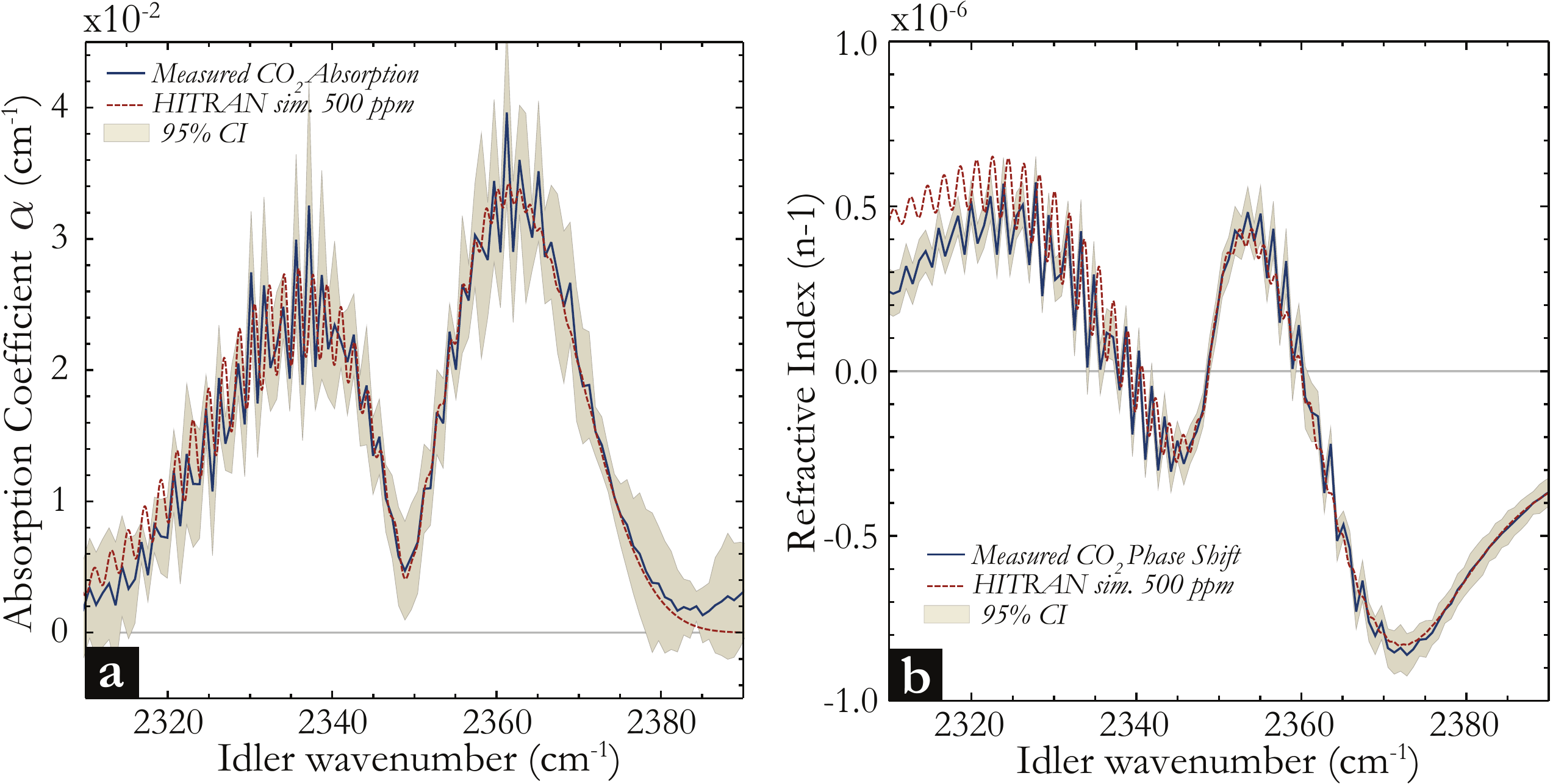}
  \caption{{\bf a.} The reconstructed absorption coefficient (blue line) of CO$_2$ with a theoretical comparison (red dashed line) to a {\it HITRAN} simulation of an atmospheric CO$_2$ concentration of \SI{500}{ppm} with a specified device resolution of \SI{1.4}{\invcm}. The beige shaded region specifies the 95\% confidence interval of the reconstructed absorption coefficient. {\bf b.}~The reconstructed refractive index (blue line) with the corresponding 95\% confidence interval. The theoretical refractive index (red, dashed) was obtained via a Hilbert transform of the {\it HITRAN} simulation presented in a.}
  \label{fig:CO2Figure}
\end{figure*}

The measured absorption coefficient for CO$_2$ is presented in Fig.~\ref{fig:CO2Figure}~a, alongside a HITRAN simulation of the absorption coefficient of CO$_2$ at an assumed concentration of \SI{500}{ppm} for ambient lab air and an instrument resolution of \SI{1.5}{\invcm} to match that of the spectrometer. The theoretical and experimental measurements show good agreement. Alongside the Boltzmann distribution which describes the envelope of the absorption feature, the fine structure that underlies it is also evident. Using the relative phase obtained from the initial fitting routine, and subsequently fitting the increase in phase shift with increasing path length, the refractive index was also reconstructed (Fig.~\ref{fig:CO2Figure}~b). Here, we compare the reconstructed phase (blue line) with the theoretical simulation obtained via the Kramers-Kronig transformation of HITRAN simulation of presented in Fig.~\ref{fig:CO2Figure} a (red line). The fine structure of the CO$_2$ anti-symmetric stretching is also clearly visible in the reconstructed refractive index. The deviation from the theoretical curves apparent around \SI{2300}{\invcm} is attributed to the sharp rise in absorption of KTP beyond \SI{4.3}{\um}.

\section*{Discussion}

We have presented two distinct approaches to the task of mid-IR spectroscopy with undetected photons. Furthermore, as the setup is interferometric, both the real and complex parts of the refractive index can be reconstructed. One can envisage other approaches to extracting the sample information from the measured signal spectra - such as a constrained fit of the measured signal spectrum, potentially at small interferometric displacements. Here though, the dispersion experienced by the biphoton from both the setup and the sample itself, introduces ambiguities in the measured signal - for instance, what should be attributed to loss, and what to destructive interference. At larger interferometric displacements these effects are decoupled, with the existing dispersion simply resulting in a frequency chirp across the spectrum. This allows for the independent extraction of both the instantaneous phase (and frequency) and the instantaneous amplitude in a single shot technique.

In our measurements we typically observe an interference visibility of between 30-\SI{40}{\percent} without a sample present. Even when accounting for losses from optical interfaces and components, the anticipated visibility should be higher (especially as the visibility scales with the square of the transmission). We attribute this uniform additional degradation to residual spatial distinguishability. Further improvements to the visibility are likely accessible by an improved experimental layout, and optimised focusing into the crystal. These optimisations should also increase the brightness and thus illumination on the CCD, providing further improvements to the signal-to-noise ratio.

While in all our measurements we have used an integration time of \SI{1}{s} per spectrum, faster detection is nevertheless possible and a clear potential advantage of this CCD grating-spectrometer based approach. As long as the dark noise and the read-out noise of the spectrometer are much lower than the signal, one can favourably trade-off measurement speed for SNR with the relative shot-noise from the spectrometer sensor increasing proportional to $1 / \sqrt{t_M}$ - e.g. a 10 times faster measurement would increase the relative shot-noise only by about a factor of 3. Note that this is a much more favorable scaling condition than that of the dark-count(current)-limited detectors common in the mid-IR region. This opens up the possibility for mid-IR spectroscopy limited only by the CCD-sensor speed (assuming sufficiently bright SPDC). 

With this proof of principle demonstration we have obtained resolutions comparable to commercially available FTIR systems. In particular, the single-shot nature of the envelope approach combined with the use of high-efficiency CCD grating spectrometers facilitates acquisition speeds largely unattainable with mature mid-IR spectrometer technologies. With brighter SPDC sources, accessible for example via the significantly increased efficiency of waveguided SPDC or the use of a pump enhancement cavity, acquisition rates in the \SI{100}{kHz} are anticipated. This unique combination of acquisition speed over a large sampling bandwidth, paired with moderate resolution is potentially a powerful new tool for the mid-IR, providing a new path to hyper-spectral imaging. Equally, with access to higher resolution grating spectrometers fast, high-resolution spectroscopy will be realisable. We estimate, that with optimised grating spectrometers (4000 pixel sensor array, fully \SI{0.2}{\invcm} resolved range of \SI{50}{nm}) resolutions below \SI{1}{\invcm} would be realistically achievable in real-world, industrial applications.

\section*{Conclusion}
In conclusion, we have presented a new approach to spectroscopy with undetected photons, utilising the maturity of CCD-based near-IR grating spectrometers combined with highly non-degenerate, broadband photon-pairs to realise spectroscopy in the mid-IR. In our first proof-of-concept implementation we demonstrate high resolution (down to \SI{1.5}{\invcm}) and short acquisition times (down to 1s) across a large spectral bandwidth between \SI{3.2}{\um} to \SI{4.4}{\um}, which we all expect to be improved on by future build-for-purpose, application-ready demonstrators. With our experimental setup we have presented two distinct approaches for the task of analysis: a single shot approach emphasising speed and a scanning approach that accesses maximum resolution. This first demonstration outlines a pathway for fast, high-resolution spectrometry in the mid-IR, that, when combined with its reduced spatial and financial footprint, could further drive the real-world impact of mid-IR spectroscopy.

\bibliographystyle{ieeetr}
\bibliography{oe-spectroscopy.bib}

\end{document}